\documentclass[12pt]{article}
\usepackage{amsmath}
\usepackage{amssymb}
\usepackage{epsfig}
\usepackage{cite}
\usepackage{color,colordvi}

\begin{document}
\vspace{0.01cm}
\begin{center}
{\Large\bf   Clocks, Algebras and Cosmology} 

\end{center}

\vspace{0.1cm}

\begin{center}

{\bf Cesar Gomez}$^{a}$\footnote{cesar.gomez@uam.es}

\vspace{.6truecm}

{\em $^a$
Instituto de F\'{\i}sica Te\'orica UAM-CSIC\\
Universidad Aut\'onoma de Madrid,
Cantoblanco, 28049 Madrid, Spain}\\

\end{center}


\begin{abstract}
\noindent  
 
{\small

}

\end{abstract}
Gauge invariant local observables describing primordial scalar quantum fluctuations in Inflationary Cosmology are identified as elements of a type $II$ de Sitter crossed product algebra. This algebra is defined, after adding a reference frame clock, as the algebra of clock dressed local operators. Clock dressing sets, in the weak gravity limit, the Schrodinger equation for gauge invariant quantum fluctuations. Instead of using a slow roll inflaton potential to define the clock Hamiltonian and the clock state we suggest a natural double de Sitter clock making the whole algebra associated with the planar patch a type $II$ factor. The corresponding clock states are EPR squeezed states. Using this clock to define the needed clock dressing leads to concrete model independent predictions of the inflationary parameters. Some speculative remarks on potential ways to define a type $I$ upgrading are briefly discussed.

\thispagestyle{empty}
\clearpage
\tableofcontents
\newpage
\section{Introduction}
The problem of time  in Quantum Gravity is an old and extensively discussed problem along the years. In a nutshell the problem is that in QG physical local observables should commute with the Hamiltonian i.e.  $[H,{\cal{O}}]=0$ and therefore they are stationary. This, as originally pointed out by D.Page and K.Wootters \cite{Page}, is similar to the problem with superselection charges where physical local observables commute with the charge generator implying that quantum superpositions of states with different charge are unobservable. A natural way to account for the non trivial time dependence we  observe in the world is by either adding an external reference frame/ observer/clock, in the sense originally suggested by Aharonov and Susskind \cite{AS}. Recently this approach has acquired a new impulse thanks to the work of E.Witten and collaborators \cite{Witten1,Witten2,Witten3,Witten4,Witten5} on the von Neumann algebras of local observables for local quantum field theory in a fixed background with horizons \footnote{The notion of reference frames to define superselection charges appears in quantum information theory \cite{RF}. The interpretation of clocks as the reference frame needed to define gauge invariant observables in QG was discussed in \cite{Gomez0} and more recently in \cite{Susskind1}.}.

The problem of how to define a clock, or more precisely an hermitian operator telling time, goes back to the early days of quantum mechanics. More specifically to the discussions on how to interpret the uncertainty relation
\begin{equation}\label{uncertainty}
\Delta E \Delta t \geq \hbar
\end{equation}
The simplest interpretation of (\ref{uncertainty}) is to consider a quantum wave packet of width $\Delta(E)$ and to interpret $\Delta t$ as the time during which the wave packet does not change significantly. In this interpretation (\ref{uncertainty}) follows from the standard Fourier analysis and simply tells us about a typical life time of the system.  As originally observed by Mandelstam and Tamm \cite{MT} in this approach we use as clock some observable  $A$ of the system that varies with time i.e.
\begin{equation}
[H,A] = \hbar \dot A
\end{equation}
for $H$ the Hamiltonian of the system and we select a quantum state on which the average $\langle \dot A \rangle$ is not changing significantly and such that the uncertainty in $\dot A$ satisfies  $\frac{\Delta \dot A}{\langle \dot A \rangle} << 1$. In this case we easily get on this state that 
\begin{equation}
\Delta t \Delta H \geq \hbar
\end{equation}
for $\Delta t \equiv \frac{\Delta A}{\langle \dot A \rangle}$. This interpretation is, however, not making any reference to any underlying measurement process.

The other popular interpretation of (\ref{uncertainty}) is as implying that energy measurements cannot be carried out in arbitrary short periods of time. In order to substantiate this interpretation we need to have some physical procedure to  identify {\it the time at which the measurement is performed}. It is in order to define this measurement time that we will need an external clock \cite{AB2}.

This will be also the case whenever we intend to implement, in local quantum field theory, invariance of physical amplitudes under coordinate time reparametrizations. In such a case we need to define the reference {\it clock time} at which the local operators involved in the amplitude are acting. This will means that local observables should depend not only on the {\it coordinate time} $t$ but also on the {\it clock coordinate} used to define the way the clock, as an independent quantum system, tells us the measurement time. These {\it clock dressed operators} are part of an extended algebra and they act on the extended Hilbert space representing the observed system and the external clock. Now gauge invariance under coordinate time reparametrizations can be implemented requiring these dressed operators to commute with $H+H_c$ with $H$ representing the observed system Hamiltonian and with $H_c$ representing the clock Hamiltonian. This algebraic structure defines, in certain cases, {\it a crossed product algebra} \cite{Witten1,Witten2,Witten3} \footnote{Note that both $H$ and $H_c$ define automorphisms i.e. coordinate time $t$ translations.}.

The key observation in \cite{Witten1,Witten2} is that for systems characterized by a type $III_1$ factor while the algebra of gauge invariant local observables is trivial ( only the identity ) once you include an external clock you can transform this type $III$ factor into a type $II$ factor and to identify a non trivial set of gauge invariant local observables. Moreover in the Hilbert space of the so defined extended system you have finite density matrix mixed states with well defined von Neumann entropy that can be interpreted as Bekenstein generalized entropy \cite{Bek}. The deep ingredient of such construction does not lies in defining an operative way to {\it tell time} using the clock but on how adding the clock you manage to define a non trivial set of gauge invariant local observables.

 As stressed in \cite{Gomez0} the need of an external clock to define non trivial gauge invariant local observables is potentially crucial in understanding primordial Inflationary Cosmology. In this case gauge invariant primordial quantum fluctuations become naturally part of the crossed product of the de Sitter type $III$ factor and a {\it clock algebra} with the {\it inflaton} playing the role of clock coordinate. We will see how the gauge invariant Mukhanov-Sasaki variables \cite{Mukhanov} are a canonical example of {\it clock dressed} operators. In this approach the crucial cosmological question of Inflation is how to define the {\it right clock Hamiltonian}. We will suggest to use as clock the analog of a free "particle" moving in the planar patch. The {\it clock Hamiltonian} generating, for the so defined clock, translations in the conformal time coordinate, is given by the pure de Sitter Bogolyubov transformation for conformal time translations on the planar patch.  This Bogolyubov transformation defines a natural automorphism on the algebra describing {\it a clock in dS}. However since, when doing Cosmology we are working on the whole planar patch, containing the static patch as well as the region beyond the cosmological horizon, we naturally expect a "two mode" clock algebra ${\cal{A}}_{c} \times \tilde{{\cal{A}}_{c}}$ with each clock algebra making the algebra of gauge invariant observables, as well as its commutant, a type $II$ factor. The Bogolyubov transformation defines a natural automorphism of this "two mode" clock algebra. Moreover the natural {\it clock states} are squeezed EPR entangled two modes states. In this frame the non trivial and observable features of Inflationary Cosmology will appear when we consider amplitudes of operators clock dressed with this {\it two mode dS-clock} \footnote{The need in Inflation of two clocks is physically quite natural. We need to define, using clock dressing, the local gauge invariant quantum fluctuations describing the CMB postinflationary correlators as well as the gauge invariant primordial quantum fluctuations during inflation. A basic assumption in Inflation is that postinflationary correlators can be expressed in terms of primordial correlators at horizon exit.}

\section{The algebraic picture of de Sitter}
In this first section we will review some aspects of the algebraic approach to weakly coupled QFT in a fixed de Sitter background \cite{Witten2}. Let us denote ${\cal{H}}$ the QFT Hilbert space. Now let us consider an observer with a static patch as her causal domain and let us define ${\cal{A}}_{dS}$ the algebra of local observables that this observer can measure performing local experiments \footnote{See \cite{Witten4} for a recent discussion and precise definitions.}. The first key assumption is that it exists a representation of ${\cal{A}}_{dS}$ in the algebra ${\cal{B}}({\cal{H}})$ of bounded operators such that the so defined sub algebra of ${\cal{B}}({\cal{H}})$ is a von Neumann factor. This means: i) that ${\cal{A}}_{dS}$ = ${\cal{A}}^{''}_{dS}$ for ${\cal{A}}'_{dS}$ the commutant of ${\cal{A}}_{dS}$ \footnote{This is defined as the set of bounded operators $x$ in ${\cal{B}}({\cal{H}})$ such that they commute with all elements in ${\cal{A}}_{dS}$.} and ii) that the center of ${\cal{A}}_{dS}$ is trivial. The GNS representation of ${\cal{A}}_{dS}$ defines a Hilbert space ${\cal{H}}_{GNS}$  with states $|a\rangle$ associated to the elements in ${\cal{A}}_{dS}$ with scalar product defined by a linear form $f:{\cal{A}}_{dS} \rightarrow R$ with $\langle a|b\rangle = f(a^+b)$\footnote{More precisely each element in the GNS Hilbert space is associated with an equivalence class defined relative to ideal kernel of the linear form $f$.}. This representation is characterized by a "vacuum" defined as the state in ${\cal{H}}_{GNS}$ associated with the identity. We will denote this state $\Psi_{dS}$. Formally identifying ${\cal{H}}_{GNS}$ with the QFT Hilbert space ${\cal{H}}$ we can associate $\Psi_{dS}$ with the globally de Sitter invariant Bunch-Davis vacuum \cite{BD} in ${\cal{H}}$. 

General arguments imply that ${\cal{A}}_{dS}$ is a type $III_1$ factor. This in particular means that it exists a state dependent modular Hamiltonian $\hat h_{\Psi_{dS}}$ generating outer automorphisms of ${\cal{A}}_{dS}$ and such that $\hat h_{\Psi_{dS}}|\Psi_{dS}\rangle =0$. Geometrically $\hat h_{\Psi_{dS}}$ is associated with a global Killing of full de Sitter, inducing inverse time translations on opposite static patches. In that sense the action of $\hat h_{\Psi_{dS}}$ on the observer static patch can be identified with the generator of time translations on that static patch. However this action cannot be identified with the action of the naive static patch Hamiltonian $H$ associated with time translation is this static patch and no action on the opposite static patch. The crucial property of ${\cal{A}}_{dS}$ as a type $III_1$ factor is that this "static patch Hamiltonian" $H$ is not defined in ${\cal{H}}_{GNS}$. This means that acting with $H$ on a generic state $|\psi\rangle$ in ${\cal{H}}_{GNS}$ we move out of the space or equivalently that the norm of $H|\psi\rangle$ is infinity. This is the direct manifestation of the divergent entanglement associated with type $III_1$ factors.

The QG "time problem" in this algebraic picture becomes specially neat \cite{Witten2}. Indeed the set of physical observables for the static patch observer should be defined by requiring invariance under time reparametrizations. Defining those by the action of $\hat h_{\Psi_{dS}}$ on the static patch we get as invariant algebra ${\cal{A}}_{dS}^{\hat h}$ which unfortunately is trivial. Note that we cannot use $H$ defined above since this operator is not defined due to the type $III_1$ divergent entanglement. 

Intuitively as discussed in \cite{Gomez0} and previously in \cite{Page} this problem is similar to the one we found with superselection charges. Indeed if we declare the algebra of physical observables to be defined by $[Q,a]=0$ for some SS charge $Q$ we cannot find, in the so defined algebra, local physical operators creating any non vanishing charge and consequently superpositions of states with different charges are unobservable. In the case of using $\hat h$ instead of the SS charge $Q$ we cannot find quantum superpositions of states with different "modular energy". The natural way to address this problem in the case of SS charges is, following \cite{AS}, the introduction of an external {\it reference frame} system \cite{RF} that in the case of time reparametrizations means {\it adding an external clock}. 

Thus the suggestion in \cite{Witten2} is to define the algebra of physical local observables  i.e. invariant under time reparametrizations of the observer static patch, as
\begin{equation}\label{crossed}
{\cal{A}}^{cr} = ({\cal{A}}_{dS} \otimes {\cal{A}}_{clock})^{(\hat h + \hat H_{c})}
\end{equation}
provided we can define a clock algebra ${\cal{A}}_{clock}$ and a clock Hamiltonian $\hat H_{c}$. Before discussing the nature of this crossed product algebra let us pause to discuss how to define a physical clock in quantum mechanics.

\section{Clock Algebra}
Following \cite{AB2} we will define a generic quantum mechanical {\it clock} as follows. Let us first define the clock as {\it an independent quantum system}. In the simplest case we can define this system as having just one degree of freedom. Thus quantum mechanically this clock is characterised by position $\hat z$ and momentum  $\hat p_z$ operators satisfying 
\begin{equation}\label{equ0}
[\hat p_z,\hat z] =i \hbar
\end{equation}
To use this system as a clock i.e. to use the value of the coordinate $z$ to tell us the clock time, we need to define its  Hamiltonian $\hat H_c$ and to define $\dot{\hat z}$ by the corresponding Heisenberg equation of motion $i\hbar \dot {\hat z} =[\hat H_c,\hat z]$. In the simplest illustrative case we can use the free Hamiltonian $\hat H_c= \frac{\hat p_z ^2}{2M}$ and to use as a clock a free particle that will tell us the time by its position. In this discussion  we consider the clock, not as a classical external apparatus, but as an external purely quantum system. Obviously to use this quantum mechanical system as a clock we need to consider a clock quantum state $|\Phi_{clock}\rangle$ for which the variance of $\dot{\hat z}$ defined as $\frac{\Delta \dot {\hat z}}{|\langle  \dot {\hat z} \rangle|}$ is negligible. Let us now ask ourselves under what conditions we can define, associated with this quantum system, a clock algebra formally defined by two self adjoint operators $\hat H_c$ and $\hat t$ satisfying
\begin{equation}
[\hat H_c,\hat t]= i\hbar
\end{equation}
For the simplest case of $\hat H_c$ the free Hamiltonian we can define
the hermitian time operator $\hat t$ as \cite{AB2}
\begin{equation}\label{equ1}
\hat t = \frac{1}{2}M(\hat z \frac{1}{\hat p_z} + \frac{1}{\hat p_z} \hat z)
\end{equation}
that indeed formally leads to 
\begin{equation}\label{equ2}
[\hat H_c,\hat t] = i\hbar
\end{equation}
and to the standard Heisenberg uncertainty relation $\Delta(\hat t)\Delta (\hat H_c) \geq \hbar$.

Obviously the previous definition of $\hat t$ should be wrong. Indeed the problem of defining a self adjoint time operator $\hat t$ canonically conjugated to some Hamiltonian is that in such a case both $\hat t$ and $\hat H_c$ should have as spectrum the whole real line. This cannot be the case if the clock Hamiltonian is positive.  In the previous simple example, where we used a free particle Hamiltonian, this problem is revealed by the fact that the operator $\hat t$ defined by (\ref{equ1}) is not well defined for $p_z=0$. 

The Hilbert space representation of (\ref{equ0}), describing the physical system used as clock, is $L^2(R)$, however the {\it clock time operator} $\hat t$ given by (\ref{equ1}) is, in principle, only well defined in the subspace of states satisfying $|\langle\psi|\frac{p_z}{M}|\psi\rangle| > 0$ i.e. only on clock states with some non vanishing $\dot z$.

In order to make a preliminary contact with the discussion in \cite{Witten2} notice that we have, in principle, two alternative ways to define the {\it external} clock. In the case presented above you start with a quantum system characterised by the algebra (\ref{equ0}) and for a given {\it positive Hamiltonian} $\hat H_c$ you try to define a clock time operator $\hat t$ satisfying (\ref{equ2}). This as discussed is not possible for   $\hat H_c$ bounded. Hence you need to define $\hat t$ on an appropriated subspace characterised, in the case of using as clock a free particle, by {\it the constraint} $|\langle \dot {\hat z} \rangle | > 0$. However, you can also start with the algebra  (\ref{equ2}) that implies the full real line as spectrum of $\hat H_c$ and try to implement {\it the positivity constraint} projecting on states with positive eigenvalue of $\hat H_c$.
Thus, in one version, in which you keep track of the physical degrees of freedom defining the external clock, you need to introduce the constraint $|\langle \dot {\hat z} \rangle | > 0$ while in the case you directly start with the algebra (\ref{equ2}) you need to introduce the positivity constraint on $\hat H_c$. This is the constraint that in the case of de Sitter leads to a type $II_1$ crossed product by contrast to a type $II_{\infty}$ factor \cite{Witten2}. 

More specifically if we define the algebra ${\cal{A}}_{clock}$ by (\ref{equ2}) then, in the case of de Sitter, you need to implement the positivity constraint on $\hat H_c$ at the algebraic level as
\begin{equation}\label{Posit}
\tilde {\cal{A}} = \Pi {\cal{A}}^{cr} \Pi
\end{equation}
with $\Pi$ the positivity projector and  ${\cal{A}}^{cr}$ the crossed product (\ref{crossed}).

The previous problem can be presented, in simpler terms, as the impossibility to define, in quantum mechanics, an ideal clock with positive Hamiltonian and {\it vanishing probability amplitude to move backwards in time} \cite{UW} \footnote{The essence of the problem is that for a bounded clock Hamiltonian if we impose that the clock goes always forward or backwards in time then general analyticity arguments based on the positivity of the clock Hamiltonian will imply that the clock is not working at all.} .

\subsection{The quantum mechanical picture}
Let us momentarily ignore this problem and consider first generic quantum states of the clock. These, for a free particle clock, are characterised by wave packets
\begin{equation}
\Phi(z,t) = \int dp_z f(p_z) e^{\frac{i}{\hbar}( p_z z -\frac{p_z^2 t}{2M})}
\end{equation}
To define a reasonable clock it is natural to use for $\Phi$ a narrow wave packet in coordinate $z$ and to use a large value of $M$ to be sure that the wave packet spreads slowly. Let us denote this wave packet $\Phi_0$. 

If we intend to use this clock to tell at what clock time the observed system has a certain property ${\cal{O}}$ we need to 
associate with the measurement of ${\cal{O}}$ a {\it measurement time} and to read such measurement time in our clock using the clock coordinate $z$. In other words, when the {\it observer} measures the observable ${\cal{O}}$ she needs to identify at what clock time as indicated by the value of $z$ this measurement was performed. In standard quantum mechanics this means that the wave function of the observed system on which you measure ${\cal{O}}$ will depend not only on the degrees of freedom $x$ of the observed system and on the particular way the observable ${\cal{O}}$ is acting but also on the degrees of freedom $z$ of the clock used to define the {\it measurement time}. Thus if we consider a clock characterised by a wave packet in $z$ coordinates that spreads slowly the wave function describing the observed system after acting with the observable ${\cal{O}}$ {\it at the clock time indicated by $z$} will be a {\it product state} of the type
\begin{equation}\label{product}
\Psi_{{\cal{O}}}= \Phi_0(z,t) \psi_{{\cal{O}}} (x,z,t)
\end{equation}
with $\Phi_0$ the clock wave packet and with $\psi_{{\cal{O}}} (x,z,t)$ representing the wave function of the observed system after acting with the observable ${\cal{O}}$ at clock time characterised by the clock coordinate $z$.

The Schrodinger equation for $\Psi$ is
\begin{equation}
i\hbar \frac{d}{dt} \Psi = ( \hat H_c + \hat H) \Psi
\end{equation}
where  $\hat H$ is the Hamiltonian of the observed system. Assuming (\ref{product}) and slow spread of the clock wave packet i.e.
\begin{equation}
\Phi_0(z,t) = f(z-\dot z t) e^{\frac{i}{\hbar}( \bar p_z z -\frac{\bar p_z^2 t}{2M})}
\end{equation}
with $\bar p_z$ the average value and $\dot z= \frac{\bar p_z}{M}$,
we get for $\psi_{{\cal{O}}}$ the Schrodinger equation
\begin{equation}\label{main}
i\hbar \frac{d}{dt} \psi_{{\cal{O}}} = (  \hat H - i\hbar \dot z \frac{d}{dz} ) \psi_{{\cal{O}}}
\end{equation}

\subsection{Clock dressing}
Formally for any local operator ${\cal{O}}$ and any {\it initial} product state $\Psi = \Phi_0 \psi_0$ in the extended Hilbert space describing the observed system and the physical clock, we can define the action of ${\cal{O}}$ on the extended Hilbert space as
\begin{equation}\label{defO}
\hat {\cal{O}} \Psi = \Psi_{{\cal{O}}} =  \Phi_0(z,t) \psi_{\cal{O}}(x,z,t)
\end{equation}
for $\psi_{{\cal{O}}}$ satisfying (\ref{main}). We can think of the so defined $\hat {\cal{O}}$ as the {\it clock dressed} operator.

Intuitively for an observer equipped with an algebra ${\cal{A}}$ of local observables once we want {\it to tell at what clock time the observed system has the properties described by measuring the observables in ${\cal{A}}$} she needs to extend the algebra by adding the clock algebra and to dress the operators ${\cal{O}}$ in ${\cal{A}}$ appropriately, for instance defining the former $\hat{\cal{O}}$. Note that in case we impose invariance under time coordinate reparametrizations the clock frame and the corresponding dressing is compulsory in order to define physical amplitudes.

The simplest way to think about {\it clock dressing} is to imagine the observable ${\cal{O}}$ as a local operator and to identify the representation of measuring ${\cal{O}}$ at certain {\it coordinate time}  $t_0$ as the smearing of ${\cal{O}}$ with some test function with support in coordinate time in a small region around $t_0$. However, {\it invariance under reparametrizations of the time coordinate} will force that operator to commute with the Hamiltonian $\hat H$ of the system we are observing and eventually, as it happens in the case of de Sitter where this Hamiltonian is the modular Hamiltonian of a type $III_1$ factor, to be just the identity operator. Nevertheless if we use a {\it quantum clock} to tell at what clock time the measurement of ${\cal{O}}$ occurred, we could do it dressing the operator and making it to depend on the clock coordinate ($z$ in the previous example ). This dressed local operator, let us say $\hat {\cal{O}}(t,z)$, depends on {\it coordinate time} $t$ and on {\it clock time} $z$. In case we now impose invariance under {\it coordinate time reparametrizations} we need to require 
\begin{equation}\label{constraint1}
[\hat H+\hat H_c,\hat {\cal{O}}]=0
\end{equation}

The dressed operator that automatically satisfies this condition is
\begin{equation}\label{timeope}
\hat {\cal{O}} = e^{i\hat H\hat t}{\cal{O}}e^{-i\hat H\hat t}
\end{equation}
for $\hat H$ the Hamiltonian of the observed system and {\it $\hat t$ the clock time operator satisfying $[\hat H_c,\hat t]= i\hbar$}.

Intuitively the operator (\ref{timeope}) acting on the extended Hilbert space of the observed system and the external clock sets the {\it measurement} of ${\cal{O}}$ at the clock time {\it measured} by $\hat t$. 

Note that the time operator $\hat t$, conjugated to the clock Hamiltonian $\hat H_c$, naturally appears when we try to solve the equation (\ref{constraint1}) defining the set of {\it gauge invariant} physical observables.
The dressing defined by (\ref{timeope}) is the typical dressing defining the de Sitter crossed product algebra in case we use for ${\cal{O}}$ operators in ${\cal{A}}_{dS}$. The dressing defined in (\ref{defO}) agrees with that defined in (\ref{timeope}) if
\begin{equation}
i\hbar \frac{d}{dt} \Psi_{{\cal{O}}} =0
\end{equation}
This exemplifies that the clock dressing agrees with the crossed product dressing once we impose invariance under  reparametrizations of the coordinate time \footnote{
Recently in \cite{Susskind1} the relation between clock dressing and time reversal is discussed. In essence the problem is that generic correlators $\langle A(t_1)A(t_2)\rangle$ only make sense, if we impose invariance under time reparametrizations, for the corresponding clock dressed operators. As discussed these dressed operators can be formally thought as depending on {\it clock time} $z$. For a clock state $\Phi_0$ the way the clock move forward or backwards in clock time depends on the sign of $\dot z=\frac{\bar  p_z}{M}$. The observation in \cite{Susskind1} is that once we choose a clock state $\Phi_0$ with a definite sign of $\dot z$ we fix time reversal {\it gauge invariance}. The interesting aspect for us is that, as stressed above, $\dot z=0$ should be projected out in order to have a well defined $\hat t$ operator.}. The important lesson to be extracted from the previous exercise is that in order to define gauge invariant observables with respect to time coordinate reparametrizations we need i) to add a quantum system working as a clock and characterised by a clock Hamiltonian ii) to define the operator $\hat t$ conjugated to the clock Hamiltonian and iii) to implement the positivity constraint on the spectrum of the clock Hamiltonian. 

\subsection{A comment on gravitational limits to clock dressing}
In the previous simple discussion of clock dressing in quantum mechanics we were using as the clock needed to define the {\it measurement time } a simple system, namely a free moving particle that tells us time by its position. Implicitly we were assuming \cite{AB2} that the mass $M$ of this external clock was large enough. Indeed equation (\ref{main}) was written in the limit where we ignore corrections of order $\frac{1}{M}$. Physically this condition on the clock mass $M$ can be easily understood once we use the clock to define dressed local operators. If we want to give meaning to the measurement time, as defined by the clock, associated to acting with a local observable ${\cal{O}}(x)$ we need to assume that the action of the operator ${\cal{O}}$ on the observed system is taking place in a region of space {\it smaller} than the typical {\it wave length} of the clock. Only in that case we can define a dressed operator that inform us about the value of the observable ${\cal{O}}$ at some concrete clock time. In more concrete terms this means that the support of the smearing function used to define the local observable should be larger than the clock wave length. Once this condition is satisfied we can, in particular, ignore $\frac{1}{M}$ corrections to the Schrodinger equation for the clock dressed operator. In case we work in the limit of $G_N=0$ this condition can be easily satisfied since we don't have a priori any limitation on the clock mass $M$. However the situation changes when gravity is turned on i.e. when we work with $G_N$ non vanishing. 

In this case the clock mass cannot be larger than $M_P$ if we don't want to have as clock a black hole. This makes clear that the $\frac{1}{M}$ corrections to the Schrodinger equation for the clock dressed operators become, once gravity is turned on, quantum gravity corrections. Hence the picture we get is the following. Once we impose general invariance under time coordinate reparametrizations we need to dress our operators which in practice means to make them dependent on the {\it clock coordinate}. The effect of the clock Hamiltonian on the amplitudes for clock dressed operators defines, depending on the particular clock state, a power series in the clock mass. Moreover the clock mass is restricted by the condition that the clock should not become a black hole. Obviously this gravitational limit on the clock mass implies also a limit on the clock precision, in particular on the clock wave packet. In particular the clock dressed operators defining the gauge invariant observables i.e. the ones invariant under time coordinate reparametrizations, should be defined using smearing functions with support larger than the clock wave length. In what follows we will work in the weak gravity limit, however we will make some general comments on the effects of clock dressing when $G_N$ is turned on in the last section. 

\section{dS maximal entropy state}
Assuming that the algebra (\ref{Posit}) for de Sitter is a type $II$ factor implies the existence of a trace $tr$ on $\tilde {\cal{A}}$ as well as the existence, for any state $|\hat \Phi\rangle$ in the extended Hilbert space representation of $\tilde {\cal{A}}$, of a density matrix operator $\rho_{\hat \Phi}$ such that for any $\hat a \in \tilde {\cal{A}}$ we have
\begin{equation}\label{definition}
\langle \hat \Phi|\hat a|\hat \Phi\rangle = tr(\rho_{\hat\Phi} \hat a)
\end{equation}
Let us consider the state $|\hat \Psi_{dS}\rangle$ representing the QFT vacuum in fixed pure de Sitter and a clock in thermal equilibrium at temperature $\beta_{dS} = r_{dS}$\footnote{Recall that thermal states in dS are formally defined by Euclidean continuation where the Killing vector associated to the modular Hamiltonian $\hat h_{dS}$ defines translation on a circle. Before adding the clock and extending the algebra the Bunch Davis vacuum $|\Psi_{dS}\rangle$ is not associated to any density matrix. By a clock in thermal equilibrium we will mean a clock state defined by a simple superposition of eigenstates of the clock Hamiltonian defined by a formal thermal probability distribution $f(\epsilon)=e^{-\beta_{dS}\epsilon}$. The meaning of this state will become clear once we define a trace form on the algebra \cite{Witten1}.} . Let us now assume that this state is in the extended Hilbert space representing $\tilde {\cal{A}}$ and therefore that we can associate with it a density matrix operator $\rho_{\hat\Psi_{dS}}$. We expect that the von Neumann entropy $s_{vN}(\rho_{\hat\Psi_{dS}}) =tr(\rho_{\hat\Psi_{dS}} \ln \rho_{\hat\Psi_{dS}})$ defined for this density matrix will be the Gibbons Hawking (GH) entropy \cite{GH} of pure de Sitter since the external added clock is in thermal equilibrium. The novelty, relative to Gibbons and Hawking, is that now we are using the extended algebra $\tilde{\cal{A}}$, including the clock, in order to have a well defined density matrix and a well defined von Neumann entropy. In GR the GH entropy is given by $\frac{A}{4G_N}$ for $A$ the area of the cosmological horizon. Thus we need to discover the relation between this GH entropy and the vN entropy $s_{vN}(\rho_{\hat\Psi_{dS}})$.  In the QFT weak gravity $G_N=0$ limit {\it where we are defining} $\tilde{\cal{A}}$ and the corresponding vN entropy, this GR entropy is infinity. For $\tilde {\cal{A}}$ a type $II_1$ factor we can {\it define} a finite vN entropy by formally substracting this infinity. Indeed we can identify $\rho_{\hat\Psi_{dS}}$ with the identity in $\tilde{\cal{A}}$ and to normalize $tr1=1$. If we do that we get from (\ref{definition}) that for any $\hat a$ in $\tilde{\cal{A}}$ we have
\begin{equation}\label{trace}
tr\hat a= tr(\hat a \rho_{\hat\Psi_{dS}}) = \langle \hat\Psi_{dS}|\hat a |\hat\Psi_{dS}\rangle
\end{equation}
Assuming $\hat\Psi_{dS}$ is a product state $\Psi_{dS} \Phi_{clock}$ with $\Psi_{dS}$ the Bunch Davis vacuum and $\Phi_{clock}$ representing the clock in thermal equilibrium i.e. defined by a thermal distribution $e^{-\beta \epsilon}$ on the spectrum $\epsilon$ of $\hat H_c$ we get $\hat\Psi_{dS}= \int d{\epsilon} e^{-\frac{\beta\epsilon}{2}} \Psi_{dS} |\epsilon\rangle$. The key consistency check with the assumption that $\rho_{\hat\Psi_{dS}}$ for this product state is the identity i.e. flat entanglement, requires to show that the trace defined by (\ref{trace}) satisfies indeed the trace property. This, non trivial check, was done in \cite{Witten1} using the KMS property. The state $\hat\Psi_{dS}$ is, in the weak gravity limit, {\it the maximal entropy state} \cite{Banks1,Banks3,Banks4,Torroba}. What is here special is not that we can renormalize the $G_N=0$ limit of Gibbons Hawking entropy, what is special is that
we can fix the substracted value of the entropy as the finite trace of the identity $tr(1)$ in $\tilde{\cal{A}}$ in the case $\tilde{\cal{A}}$ is a type $II_1$ factor.
Recall that the type $II_1$ property and consequently the existence of a maximal entropy state is a consequence of imposing positivity for the clock Hamiltonian \footnote{Note that although the von Neumann entropy for $\rho_{\hat\Psi_{dS}}$ is normalized to zero with $\rho_{\hat\Psi_{dS}}$ identified with a flat entanglement density matrix we need to keep a finite $\beta_{dS}$ in the definition (\ref{trace}) of the trace. In other words what represents flat entanglement is not the non existent density matrix describing pure dS but the density matrix describing the extended system including a clock with a simple clock state characterised by the GH finite $\beta$.}. 

The clock defined in "thermal equilibrium" is a very bad clock. Indeed the time uncertainty in $\hat t$ is order $\beta$ since the typical energy uncertainty of the clock is given by the GH temperature. What about using a better clock ? Such a clock must have average energy larger than GH temperature and therefore we should expect that it will modify the de Sitter metric. To see how this is working let us define the clock quantum state as a wave packet with average value $E$ of the clock energy and such that it spreads slowly. We can characterise this wave packet as $\Phi_{clock} = \int d\epsilon f(\epsilon)|\epsilon\rangle$ for an appropriated $f$. We can now assume that the state in the extended Hilbert space representing just the presence of this clock is a product state $\hat \Phi =\Phi \Phi_{clock}$ with $\Phi$ some state in the QFT Hilbert space ${\cal{H}}$ {\it different from the BD vacuum}. The state $\hat \Phi$ will account for the back reaction of the clock energy on the background metric. The first hint about this state is to expect that the corresponding vN entropy will differ from the vN entropy of the maximal entropy state, that is normalised to zero, by an amount of the order $E\beta$. This is in particular the expected deviation of GH entropy if we assume that the metric in the presence of a localised clock of energy $E$ is a Schwarzschild-de Sitter metric. Thus we expect the vN entropy of $\rho_{\hat Phi}$ to contain a piece of the order $\langle \hat \Phi|\beta H_c|\hat \Phi\rangle$ for $H_c$ the clock Hamiltonian. As shown in \cite{Witten2} this is indeed a piece of the vN entropy associated to the density matrix $\rho_{\hat \Phi}$. This vN entropy is a Bekenstein generalized entropy \cite{Bek} and it contains two pieces. The other piece will depend on what is, in this case, the QFT state $\Phi$ and in particular how different is this state, representing now the analog of the QFT Bunch Davis vacuum but for the Schwarzschild-de Sitter background, from $\Psi_{dS}$. Assuming the clock state to spread slowly i.e. to have small time uncertainty, we can use Araki's relative operator $\Delta(\Psi_{dS}|\Phi)$ to measure the distinguishability of both states relative to the de Sitter algebra. However this general argument does not fix the state $\Phi$.

Naively we should expect that
\begin{equation}\label{constraint}
i\hbar \frac{d}{dt} |\hat \Phi\rangle = ( \hat h_{\Psi_{dS}} + \beta \hat H_c)|\hat \Phi\rangle = 0
\end{equation}
where $\hat h$ is the modular Hamiltonian. This is a naiv description of the physical Hilbert space as those states of the combined system -- that includes the external clock-- that are invariant under time reparametrizations. The important subtlety with (\ref{constraint}) is that the state $|\hat \Phi\rangle$ describes the full space i.e. not only the region associated with $\tilde{\cal{A}}$ but {\it also} the region associated with its commutant $\tilde{\cal{A}}^{'}$ and therefore it generically represents two entangled clocks \footnote{Ignoring one of the clocks will lead to some extra entropy deficit \cite{Witten2}.}.

\section{Gauge invariant observables in primordial Inflationary Cosmology}

The first ingredient needed to define the weakly coupled QFT description of quantum fluctuations during the inflationary period is to identify those scalar quantum fluctuations invariant under time reparametrizations. According to our previous discussion if the inflationary period is defined using {\it pure de Sitter}, the potential set of local gauge invariant fluctuations reduces to the invariant subalgebra ${\cal{A}}_{dS}^{\hat h}$ for $\hat h$ the modular Hamiltonian defining time translations on the observer static patch. As discussed this invariant subalgebra is, for pure de Sitter, trivial and only contains the identity operator. This means that in pure de Sitter the observer cannot identify neither any gauge invariant quantum fluctuation created in her static patch nor its potential horizon exit. Following our previous discussion the most reasonable way to get a non trivial set of gauge invariant local fluctuations describing the inflationary primordial period will be to {\it add an external clock algebra} and to define the gauge invariant quantum fluctuations using the non trivial crossed product algebra $({\cal{A}}_{dS}\times {\cal{A}}_{clock})^{(\hat h+\hat H_c)}$ or its constrained version if we include the positivity constraint on the clock energy spectrum.

Hence we expect that {\it the algebra of gauge invariant fluctuations describing the primordial inflationary phase should be a crossed product algebra for some external added clock}. This is in essence the approach discussed in \cite{Gomez0,Gomezcross}. 

In order to identify the clock let us first recall the standard definition of gauge invariant scalar fluctuations in Inflationary Cosmology. These gauge invariant scalar fluctuations are known as Mukhanov-Sasaki variables \cite{Mukhanov}. 

The most general parametrisation of linear scalar fluctuations is given by the metric
\begin{equation}\label{metric}
ds^2 = a^2((1+2\phi)d\eta^2 - 2B_i dx^{i}d\eta -
((1-2\Psi)\delta_{i,j} +2 E_{i,j})dx^{i}dx^{j})
\end{equation}
with $\phi,E,B,\Psi$ arbitrary functions of space and time and $\eta$ the conformal time related to the physical time $t$ by $t=\int a d\eta$. We are interested in identifying gauge invariant scalar fluctuations i.e. those fluctuations that are not due to pure reparametrizations of time. From (\ref{metric}) it follows that both $\phi$ and $\Psi$ change under time reparametrizations thus we need "to dress" the corresponding QFT operators, describing these fluctuations, to define gauge invariant observables. In order to define this gauge invariant dressing we could try to use a {\it clock dressing}, in the sense defined above. Mukhanov's variable \cite{Mukhanov} achieves naturally this goal. 

To see how that is working let us formally define the {\it clock coordinate} as the homogeneous inflaton vev $\varphi_0$. To define the clock algebra we also consider the canonically conjugated momentum operator $p_0$ with $[\varphi_0,p_0] = i\hbar$ and a clock Hamiltonian $\hat H_c$ such that $i\hbar\dot \varphi_0=[\varphi_0,\hat H_c]$. Thus, at this level, the external clock is defined by just one degree of freedom that we identify with the homogeneous added inflaton field and by a clock Hamiltonian $\hat H_c$.  As discussed above once we have defined this quantum mechanical system we can define, using (\ref{equ1}) and the appropriated projection, a clock time hermitian operator $\hat t$ formally satisfying $[\hat t,\hat H_c] =i\hbar$. From that it follows Heisenberg relation $\Delta t \Delta H_c \geq \hbar$. The typical uncertainty $\Delta t$ for the clock time operator $\hat t$ will depend on the particular quantum state representing the clock as $\frac{\Delta \varphi_0} {|\langle \dot \varphi_0 \rangle|}$.

 Let us now think of the metric fluctuation defined by $\Psi$ as an element $a$ in ${\cal{A}}_{dS}$ and let us look for the corresponding crossed product (clock) dressing $\hat a = e^{i\hat h \hat t}ae^{-i\hat h \hat t}$ for $\hat t$ the clock time operator. For the inflaton clock just defined we have $\hat t \sim \frac{\varphi_0}{\dot \varphi_0}$ and consequently the clock dressing of $\Psi$ becomes Mukhanov-Sasaki variable
\begin{equation}\label{variable}
\chi= \Psi +\frac{\delta \varphi_0 H}{\dot\varphi_0}
\end{equation}
for $H$ the Hubble parameter.
Thus we conclude {\it that the gauge invariant scalar fluctuations defined by MS variables are in the crossed product algebra defined for an external clock with clock coordinate $\varphi_0$ and some clock Hamiltonian $\hat H_c$ ( not yet specified )}. It is important to stress that before adding the clock we don't have any non trivial gauge invariant way to describe scalar fluctuations in pure de Sitter. Naively we could think of using the Hamiltonian $H$, generating time translations in the static patch of the observer and doing nothing on the complementary static patch, to define invariance under time reparametrizations. However, and that is a crucial consequence of the type $III_1$ nature of the de Sitter algebra ${\cal{A}}_{dS}$, this Hamiltonian is not acting on the GNS Hilbert space and consequently is not acting on the QFT Hilbert space describing the weakly coupled quantum fluctuations. Moreover using $H$ leads to undesired divergences due to the fact that $H|\psi\rangle$ for generic $|\psi\rangle$ is not normalizable.

This simple comment explains why when we {\it stop} the clock i.e. when we impose $\dot \varphi_0=0$ or equivalently pure de Sitter without clock, the physical amplitudes of gauge invariant observables become divergent and ill defined. This is specially clear for the power spectrum of scalar curvature fluctuations that becomes divergent in the no clock limit $\dot\varphi_0=0$. 

Another important comment concerns the way the clock is introduced using the homogeneous inflaton field $\varphi_0$. In this case we don't jump into adding an abstract clock algebra defined by $[\hat t,\hat H_c]=i\hbar$ but we insist on defining a quantum mechanical system with coordinate $\varphi_0$ and the corresponding conjugated momentum $p_0$ and with clock Hamiltonian $\hat H_c(\varphi_0,p_0)$ defining $\dot\varphi_0$. It is on the basis of this clock data that the clock time $\hat t$ is formally related with $\frac{\varphi_0}{\dot \varphi_0}$. 

In order to proceed as we did in the previous sections we will need to consider {\it product states} in the extended Hilbert space representing the weakly coupled QFT on fixed de Sitter background and the external clock defined using the inflaton field $\varphi_0$. As we did in section 3 let us consider the state $\Psi_{QFT;\chi} \Phi_{clock}$ obtained by acting on the extended Hilbert space with the dressed and gauge invariant MS observable $\chi$\footnote{Note that this dressed operator as defined in (\ref{variable}) belongs to the crossed product algebra.}. We can think of $\Phi_{clock}$ as some localised wave packet $\Phi(\varphi_0,t)$ but for the time being it will not be important to identify this clock state. What is crucial is to impose the gauge invariant constraint
\begin{equation}
i\hbar \frac{d}{dt}\Psi_{QFT;\chi} \Phi_{clock} = (\hat H + \hat H_c) \Psi_{QFT;\chi} \Phi_{clock}=0
\end{equation}
for $\hat H$ representing the pure de Sitter evolution. This leads, for a clock wave packet that spreads slowly, to a simple equation for $\Psi_{QFT;\chi}$ namely
\begin{equation}\label{CHM}
(\hat H + \langle\hat H_c\rangle) \Psi_{QFT;\chi} = 0
\end{equation}
for $\langle\hat H_c\rangle$ defined relative to the clock state $\Phi_{clock}$. 

What is the meaning of (\ref{CHM}) ? It is the Schrodinger equation for the gauge invariant quantum fluctuations $\chi$ {\it defined by a clock dressing} with clock Hamiltonian $\hat H_c$ and clock state $\Phi_{clock}$. This equation for $\Psi_{QFT;\chi}$ representing scalar curvature fluctuations is the well known Chibisov-Mukhanov equation \cite{CHM}. 

Before describing the connection between (\ref{CHM}) and Chibisov Mukhanov equation let us make some comments to link the present discussion with the discussion above on the effect of a clock artefact defined with a localised energy. In that case the state that now we are calling generically $\Psi_{QFT}$ was the QFT state that accounts for the back reaction on the background metric of {\it the energy of the external localised clock}. This back reacted metric can be Schwarzschild de Sitter. In such a case $\Psi_{QFT}$ was the analog of Bunch Davis vacuum {\it but} for the Schwarzschild de Sitter metric describing the clock artefact with some localised energy $E$. Now we are using as clock the inflaton vev $\varphi_0$ so the corresponding back reaction, to the presence of the inflaton clock, should be some quasi de Sitter metric. What particular quasi de Sitter metric will depend on what particular clock Hamiltonian we decide to use. Recall that to add a clock algebra is not optional if we want to have a non trivial set of gauge invariant observables. What in principle could be optional is what particular clock we decide to use and how that decision depends on some underlying physics argument.

A second relevant comment concerns the {\it tensor modes}. Those are by construction gauge invariant so they don't need to be dressed by any external clock. Algebraically we can think of these tensor modes as {\it central} with respect to the crossed product algebra.

Once that is clarified let us recall the form of Chibisov-Mukhanov equation. You define $v= z\chi$ with $z=a\frac{\dot\varphi}{H} \equiv a\sqrt{\epsilon}$. The Schrodinger equation for the Fourier mode of comoving momentum $k=0$ is
\begin{equation}
v^{''} -\frac{z^{''}}{z} v=0
\end{equation}
where $v'$ represents derivative with respect to conformal time $\eta$.
Let us rewrite $\frac{z^{''}}{z} = \frac{2}{\eta^2}(1 +\Delta)$. To make contact with (\ref{CHM}) we set the pure de Sitter part defined by $\hat H$ as $\hat H v= v^{''} -\frac{2}{\eta^2} v$ and $\frac{2\Delta}{\eta^2} $ as $\langle\hat H_c\rangle $. In other words we map the deviation $\frac{\Delta }{\eta^2}$ from pure de Sitter of Chibisov Mukhanov equation into the contribution of the average clock energy defined by the clock Hamiltonian on some particular clock state $\Phi_{clock}$. This identification assumes a non vanishing but constant $\dot\varphi$ \footnote{It is in this approximation that we can write $\frac{z^{''}}{z} = \frac{2}{\eta^2}(1 +\Delta)$ with $\Delta$ constant and time independent.}.

In this simplest approximation we have 
\begin{equation}\label{simple}
\frac{z^{''}}{z} = 2H^2a^2 + \epsilon H^2a^2
\end{equation}
The clock interpretation of this equation links $\epsilon H^2a^2$ with $\langle\hat H_c\rangle$. Thus the quasi de Sitter slow roll parameter $\epsilon$  is fully determined by the dynamics of the added clock. 

In words a non vanishing cosmological constant forces us to add a clock and to work on a Hilbert space representing the crossed product algebra. When we imagine this clock as an artefact we move, by imposing gauge invariance, into some Schwarzschild de Sitter metric while when we add the clock defined by the inflaton dynamics we naturally move into a quasi de Sitter metric. 

Let us now consider equation (\ref{simple}). The clock Hamiltonian that leads to 
$\epsilon H^2a^2$ corresponding to constant $\epsilon$ is like a clock defined by a {\it free particle} with Hamiltonian $a^2(\dot\varphi)^2$ and $\dot\varphi = \sqrt{\epsilon}H$. Thus $\epsilon$ non vanishing and constant corresponds to a "free particle clock" with non vanishing $\dot\varphi$ and vanishing "acceleration".

Once we include the space dependence of the inflaton field we can consider the different Fourier modes $\chi_k$ for $k$ the comoving momentum. Note that for each mode $k$ we are formally defining a $k$ clock with clock coordinate $\varphi_k$ and clock Hamiltonian $\hat H_{c,k}$. The Schrodinger equation for these modes becomes

\begin{equation}\label{CHM1}
v_k^{''}+ (k^2-\frac{z^{''}}{z}) v_k=0
\end{equation} 
that we can rewrite formally as
\begin{equation}\label{CHM1}
v_k^{''}+ (k^2-\frac{2}{\eta^2} -\langle \hat H_{c,k}\rangle) v_k=0
\end{equation} 
In the short wave regime where $k^2$ is much larger than $\frac{z^{''}}{z}$ the effect of the clock is effectively suppressed. After our algebraic discussion this is dangerous since once the clock is suppressed we move back into the type $III$ ill defined regime. This is a different way to present the so called {\it transplanckian problem in Cosmology}. In the long wave regime the solution to the Schrodinger equation goes as $z$ and therefore it strongly depends on the clock contribution.

Let us now consider equation (\ref{CHM1}) in the long wave limit $k\sim 0$ or more precisely $k<< \frac{z^{''}}{z}$ and let us parametrise $\langle \hat H_{c}\rangle = \frac{\Delta}{\eta^2}$ with $\Delta$ independent on time. This, as said, corresponds to a {\it constant speed clock}. In this case the solution can be written as
\begin{equation}\label{solution}
v=f(k)\frac{(\tilde k \eta)^{\delta}}{\eta}
\end{equation}
with $\tilde k$ an {\it arbitrary} ( in the $k\sim 0$ regime ) pivot scale, $\delta$ determined by the clock contribution $\Delta$ i.e. $\delta(\Delta)$ and $f$ defining the normalization. This normalization can be fixed by matching at horizon exit $k\sim aH$ with the free wave solution in the short wave limit $k^2>>\frac{z^{''}}{z}$. 

Using this normalization the power spectrum of scalar curvature fluctuations\footnote{Defined as $k^3|\chi_k(\eta)|^2$.} at horizon exit for the mode $k$ goes like
\begin{equation}\label{power}
{\cal{P}} \sim (\frac{\tilde k}{k})^{2\delta} \frac{H^2}{\epsilon}
\end{equation}
with $2\delta=(1-n_s)$ where recall $\dot\varphi = \sqrt{\epsilon}H$. This provides the first lesson. Namely in the very long wave regime the spectral index $(1-n_s)$ \cite{index,Hawking} is determined by  the {\it clock energy} $\langle \hat H_c \rangle$ once we define the clock with a constant but non vanishing $\dot\varphi$. Obviously this clock corresponding to non vanishing but constant $\epsilon$ is by no means the only possible clock neither the best candidate for the right clock. Using a more general clock makes the equation for the dressed clock gauge invariant variable (\ref{CHM1}) more complicated than the Bessel equation we have been considering above.

Now we can use the {\it time delay} interpretation of ${\cal{P}}$ as $(H\Delta(t))^2$ \cite{timedelay}. In the clock picture this should be done using for $\Delta(t)$ the {\it clock uncertainty} that is determined by $\hat H_c$ by the standard Heisenberg relation $\Delta t \Delta H_c \geq \hbar$. Thus in the approximation where we assume the clock $\dot\varphi$ non vanishing but constant we get $\Delta H_c \sim \frac{\dot\varphi \hbar}{\Delta \varphi}$ and consequently ${\cal{P}} \sim \frac{H^2 (\Delta \varphi)^2}{\dot \varphi^2}$ that leads to (\ref{power}) once we use $\Delta \varphi \sim \frac{v}{a}$ for $v$ the solution (\ref{solution}). In summary the former simple discussion can be seem as an argument in two steps. First of all we use the clock Hamiltonian and the clock state to identify the Schrodinger equation for clock dressed gauge invariant scalar fluctuations and we formally solve that equation. In the simplest case we are considering we do that in the limit of small comoving momentum $k$. If we don't impose this restriction we can try to solve the equation for arbitrary $k$ provided, in our scheme, we know the corresponding clock Hamiltonian for those modes. The solution to this dressed Schrodinger equation gives us information about the value of $\Delta \varphi$. In the second step we use an approximate time delay representation of the power spectrum as $(H\Delta(t))^2$ but for $\Delta(t)$  defined by the {\it clock time} as determined by $\varphi$. The consistency of both steps requires that the $\Delta \varphi$ determined by the dressed Schrodinger equation is precisely the one that leads to the correct time delay representation of the power spectrum as $(H\Delta(t))^2$. Note that this consistency requirement is far from trivial and as we will discuss in the next section it requires to identify a concrete clock with comoving momentum mode $k$ and with a time uncertainty that accounts for the anomalous scale dependence $(\frac{k}{\tilde k})^{2\delta}$. Above we were presenting the consistency condition for this two step argument in the simpler case where we assume a clock with $\dot \varphi$ non vanishing but not changing in time. However the previous formal argument was done without using any concrete clock Hamiltonian.

Finally let us make a comment on the potentially dangerous transplanckian regime. As already pointed out this regime acquires a different meaning in the clock dressing approach. In such a regime it looks superficially that we suppress the effect of the clock and that consequently we move back into the type $III$ regime. However we can also address this problem from the point of view of our previous discussion on {\it clock precision}. In the weak gravity limit where $G_N=0$ we do't need to be worried about increasing the "clock energy" and consequently the "clock precision".  However this changes whenever we turn on a non vanishing value of $G_N$. In this case the clock "mass" should be smaller than $M_P$ which in this algebraic approach means that local and gauge invariant scalar fluctuations should be reduced to be subplanckian. Otherwise the needed clock dressing will make the corresponding gauge invariant operators non local. This simple algebraic argument likely leads to the transplanckian bound on the maximal number of e-foldings \cite{vafa}.

The previous discussion should not come as a surprise. It simply reflects the fact that gauge invariant cosmological observables describing the quantum fluctuations of a primordial exponentially expanding phase should be thought as elements of a dS crossed product algebra i.e. as {\it clock dressed operators}. In this picture the inflaton dynamics is just defining the external clock, namely the clock coordinate as well as the clock Hamiltonian. The interesting problem is what can we do to identify a {\it natural clock} without getting lost into a {\it landscape of clocks}\footnote{A landscape of potential inflationary potentials.}. In the next section we will make a suggestion on how to identify a natural and model independent clock and how the so defined clock leads naturally to a non vanishing value for the spectral index $(1-n_s)$.

\section{The de Sitter double squeezed clock}
At this point of the discussion our problem is to identify a quantum system that can be used to define a physical clock in de Sitter and to use it to define gauge invariant quantum fluctuations through the corresponding clock dressing. Physical amplitudes for these clock dressed operators are determined by the condition of invariance under time coordinate reparametrizations. Thus the Schrodinger equation governing these gauge invariant quantum fluctuations depend, at the linear level where we are working, on the initial reference pure de Sitter background on the clock Hamiltonian and on the clock state. In principle the clock data, namely the clock Hamiltonian and the clock state can be thought as an abstract way, in case we use the inflaton vev as clock coordinate, to define a particular inflationary model potential. This will lead to an undesired richness of potential clocks to use in order to define a non trivial set of gauge invariant clock dressed observables. Thus the question we want to address in this section is if we could figure out a natural and model independent clock. 

In order to answer this question is convenient to use the planar patch of de Sitter to describe the Cosmology during the primordial inflationary period. In the planar patch we have two regions. One is the static patch of the observer and the other piece, representing the planar patch complement of the static patch, describes the part of space-time that is beyond the observer cosmological horizon. The Killing vector $\frac{\partial}{\partial t}$ in static coordinates defines opposite time translations in the two static patches, let us say $L$ and $R$, as well as opposite time translations on the two planar patch complements (the upper $U$ and lower $L$ triangles of the Penrose diagram ). Time translations in the $L$ or $R$ static patch define automorphisms of ${\cal{A}}_{dS}^{L,R}$ with the corresponding modular hamiltonian defining the automorphism on ${\cal{A}}_{dS}^{L} \times {\cal{A}}_{dS}^{R}$. Let us now define the algebras ${\cal{A}}_{dS}^{U,L}$ as the algebras of local observables with support on the region beyond the static patch cosmological horizon ( for instance the upper $U$ or lower $L$ triangle ). We could expect to define an automorphism of these algebras and the corresponding modular hamiltonian using the corresponding action of $\frac{\partial}{\partial t}$ on these regions.  In doing Cosmology we are effectively interested on the {\it planar patch alebra} ${\cal{A}}_{dS}^{L} \times {\cal{A}}_{dS}^{U}$\footnote{Or the one defined by $R$ and $L$. Note also that the action of time automorphisms on ${\cal{A}}_{dS}^{L}$ is not leading ( for any finite time) to elements of ${\cal{A}}_{dS}^{U}$. }. Let us now assume that both algebras ${\cal{A}}_{dS}^{L}$ and  ${\cal{A}}_{dS}^{U}$ are both type $III_1$ factors and let us define two type $I$ {\it clock algebras} let us say ${\cal{A}}_{c}^{L,U}$ to define the corresponding type $II$ factors. Let us formally define these two clock algebras by two clock coordinates, let us say $z$ and $\tilde z$, with the corresponding momentum operators $p_z$ and $p_{\tilde z}$ as well as two clock Hamiltonians $ H_c$ and 
$\tilde{H_c}$. 

A priori we should expect the corresponding {\it double clock state} should represent an entangled state of both "clocks". Since the planar patch can be foliated in terms of equal conformal time $\eta$ hypersurfaces the two-clock state will depend on $\eta$ as $\Phi_{clock}(\eta)$. The role of the reference pure de Sitter background we are using is to define a non trivial time evolution of the clock state reflecting the fact that the conformal time is not globally defined on the planar patch. Moreover this pure de Sitter time evolution of $\Phi_{clock}(\eta)$ is determined by the well known Bogolyubov transformation that naturally defines an automorphism on the clock product algebra ${\cal{A}}_{c}^{L}\times {\cal{A}}_{c}^{U}$. The clock state in the Hilbert space representation of ${\cal{A}}_{c}^{L}\times {\cal{A}}_{c}^{U}$ will be an entangled state, actually an EPR like entangled squeezed state \footnote{Note that we are using pure de Sitter input to define the double clock that should account for the natural clock dressing on the full planar patch.}. 

The two clock Heisenberg algebras defined by $z,p_z$ and $\tilde z,\tilde p_{z}$ leads to a squeezed state only if $[z,p_z] =i\hbar$ and $[\tilde z,\tilde p_z]= -i\hbar$. Once we include clock modes with commoving momentum $k$ and we use the standard creation annihilation operator basis both algebras correspond to $a_{k},a^{\dagger}_{k}$ and $a_{-k},a^{\dagger}_{-k}$. As said the squeezed state is generated by the Bogolyubov time transformation and it represent the {\it vacuum} for modes $k$ at conformal time $\eta$. This state $|k,\eta\rangle$ is given by (see \cite{Martin})
\begin{equation}\label{state}
|k,\eta \rangle = \sum_n c((k\eta),n) e^{in\Phi((k\eta))}|n_k,n_{-k}\rangle
\end{equation}
where we sum over all integers and where
$c((k\eta),n)=C \tanh({r(k\eta)})^{n}$
with $C$ a normalization constant, 
$r(k \eta) = - \sinh^{-1}(\frac{1}{2k\eta})$
and with the phase given by
$\Phi(k\eta) = -\frac{\pi}{4} -\frac{1}{2} \tan^{-1}(\frac{1}{2k \eta})$.
We will suggest to use this entangled and squeezed state as the natural {\it two clock entangled state} to define gauge invariant quantum scalar fluctuations on the planar patch. 

Now we need to identify for this clock the analog of the $\hat t$ operator that, as discussed we need to use to define the gauge invariant clock dressed operators. As discussed formally $\hat t$ is the conjugated to the clock Hamiltonian. For the former description of the two clock state as a squeezed state and by analogy with what is familiar for quantum coherent states we can think of $\hat t$ as conjugated to the operator number defining the two clock state energies. Thus we can map $\hat t$ with the {\it phase} $\Phi$ conjugated to the operator number and to identify the clock algebra as $[N,\Phi] = -i\hbar$ with $\Phi$ the phase in (\ref{state}).

This last comment provides the hint we need to guess how to extract from the squeezed state $|k,\eta\rangle$ the corresponding mode dependent clock time uncertainty $\Delta (\hat t_{k,\eta})$. Since all functions defining the squeezed clock state depend on the combination $k\eta$ we can extract the information about the clock time uncertainty in terms of the variance of $\frac{\partial \Phi}{\partial (k\eta)}$, that we will denote ${\cal{F}}(k\eta)$, evaluated on the squeezed state (\ref{state}):
\begin{equation}\label{tilt}
{\cal{F}}(k\eta)= ( \sum_n c(k\eta,n)^2 (\frac{\partial \Phi(k\eta,n)}{\partial (k\eta)})^2 - (\sum_n c(k\eta,n)^2 (\frac{\partial \Phi(k\eta,n)}{\partial (k\eta)}))^2)
\end{equation}
Defining $\hat {\cal{F}}(k\eta) = (k\eta)^6 {\cal{F}}(k\eta)$ we can now try to extract direct information on the power spectrum $(\frac{k}{\tilde k})^{(1-n_s)} \frac{1}{\epsilon}$ in terms of the {\it squeezed clock} time uncertainty set by ${\hat {\cal{F}}(k\eta)}$.  The numerical evaluation of $\hat {\cal{F}}(k\eta)$ leads \cite{GJ,GJ2} to
 \begin{equation}
 \hat {\cal{F}}(k\eta) \sim \frac{(k\eta)^{\alpha_F(k\eta)}}{(k\eta)^2}
  \end{equation}
for a "quantum tilt" $\alpha_F(k\eta)$ \cite{GJ}\footnote{This quantum tilt evaluated in \cite{GJ} is a well defined function of $k\eta$ that depends numerically on how we cut the sum in (\ref{tilt}) i.e. on the numerical bound on the number of entangled pairs used to evaluate ${\cal{F}}(k\eta)$. The numerical sensibility of the result on this cut was discussed in \cite{GJ}.}.
Since we want to use ${\cal{F}}(k\eta)$ to identify the power spectrum at horizon exit we should think of the argument $k\eta$ as determined by {\it the threshold condition} $k^2\sim \frac{z^{''}}{z}$ {\it but reduced to the clock correction} i.e. $k\delta(k)\sim \frac{\Delta}{\eta^2}$. This leads to two conditions, namely the one that relates $\Delta$ to some $k\eta$ as $k\eta\sim 2\Delta$ and $(1-n_s)$ to the value of $\alpha_F$ at this particular value of $k\eta$ i.e. $\alpha_F(2\Delta)= 2\delta(\Delta)$ with $\delta(\Delta)$ defined by (\ref{solution}). Thus, taking as input the numerical value of the function $\alpha_F(k\eta)$, we can extract the concrete numerical values of the inflationary slow roll parameters\footnote{As well as the pivot scale defining the CMB regime.}. 

Since our target in this note is simply to show the existence of a predictive recipe we will not insist neither on the concrete numerical values nor on the potential phenomenology delivered by this scheme already discussed in \cite{GJ,GJ2}\footnote{In the approximation used in \cite{GJ} we get a nice value of $(1-n_s) = 0.0328$.}. The key message can be summarized saying that {\it the squeezed two mode clock state $|k,\eta\rangle$ naturally delivers a clock time uncertainty depending on $(k\eta)$ and with a non trivial scale dependence on $k\eta$ encoded in $\alpha_F$}. Matching this clock uncertainty with the power spectrum for scalar perturbations at horizon exit it sets the values of the inflationary slow roll parameters as well as the typical threshold scale i.e the range of wave lengths at which $k^2\sim \frac{z^{''}}{z}$. 

The key lesson we would like to extract from this exercise is that we can use {\it pure de Sitter information} to define a two modes natural planar clock and to learn about the observable power spectrum, using, to define the needed clock dressing of cosmological observables, this de Sitter double squeezed clock. 

\section{Some speculative conjectural final remarks}
The main target of this note has been to define amplitudes for local and gauge invariant cosmological observables using a type $II$ clock dressing. In addition we have suggested the existence of a natural dS clock with clock Hamiltonian defined by the planar dS Bogolyubov transformations and with a double EPR squeezed clock state.

This construction motivates some natural questions. The first and probably the most fundamental one is to unveil how to use quantum gravity effects in dS ( recall we have been working in the weak gravity limit $G_N=0$ ) to define a potential upgrading to a type $I$ factor. Adding a reference frame clock in the $G_N=0$ limit makes the type $III$ QFT factor into a more friendly type $II$ factor. Thus what do we need "to add" to get a type $I$ factor ? The type $II$ appears because invariance under time reparametrizations forces us to add the reference frame clock. As briefly discussed this reference frame clock can have in the weak gravity limit $G_N=0$ arbitrary precision, however that is not the case once we turn on gravity. At the level of amplitudes for dressed operators this means that the root of the intrinsic locality constraints, once we include quantum gravity $G_N$ non vanishing effects, lies in the needed clock dressing \cite{hamed}. Hence, pushing into the UV gauge invariant dressed observables, comes with an unavoidable lost of clock precision i.e. the clock wave length is pushed into the IR. A conjectural way to understand this UV-IR bouncing for dressed operators could be in terms of what we can call {\it clock classicalization} \cite{us1,us2}. In other words once gravity is turned on the needed added clock flows with energy into the IR. The question then is if this {\it clock classicalization} implies effective type $I$ dynamics.

A second natural remark concerns the natural limits on the number of e-foldings that inflation lasts. A possibility briefly discussed in \cite{Gomez0} could be to link this number of e-foldings to the von Neumann-Murray dimension for the entropy deficit for the two needed clocks. Namely one for the primordial period and the other to account for clock dressing at the CMB scales.

Finally we can ask ourselves in the holographic AdS context about the meaning of the analog "clock dressing" at the dual holographic boundary. As pointed out in \cite{Liu} we expect that for Yang Mills at large $N$ and in finite volume we will have a type $III$ description of the high temperature Hagedorn phase. What is the meaning of the analog clock dressing in this non gravitational holographically dual set up ? In \cite{Gomez0} we have suggested to use the dynamics of the gapped and confined matrix model eigenvalues in the Hagedorn phase as the natural candidate to define the Yang Mills analog of the type $II$ added reference frame.

\section{Acknowledgments}
This work was supported by grants SEV-2016-0597, FPA2015-65480-P and PGC2018-095976-B-C21. Parts of this work were presented  in the Cambridge workshop {\it Quantum de Sitter Universe}. I thank the organisers for the opportunity and for the great and insightful meetting.

 \end{document}